\documentclass[twocolumn,showpacs,preprintnumbers,amsmath,amssymb]{revtex4}
\usepackage{graphicx}

\begin{document}

\title{Peculiar properties of oscillator with nonlinear coordinate-dependent mass.}

\author{B.\,I.\,Lev}          %\email{blev@bitp.kiev.ua}
\author{V.\,B.\,Tymchyshyn}   \email{yu.binkukoku@gmail.com}
\author{A.\,G.\,Zagorodny}    %\email{a.zagorodny@bitp.kiev.ua}

\affiliation{Bogolyubov Institute for Theoretical Physics,
National Academy of Sciences of Ukraine, Metrolohichna 14-b, Kyiv 03680, Ukraine.}
\date{\today}

\begin{abstract}
A nonlinear model of the scalar field with a \emph{coupling} between the field and its gradient is developed.
It is shown, that such model is suitable for the description of phase transitions accompanied by formation of spatially inhomogeneous distributions of the order parameter.
The proposed model is analogous to the mechanical nonlinear oscillator with the coordinate-dependent mass or velocity-dependent elastic module.
Besides, for some values of energy the model under consideration possesses exact analytical solution.
We assume, this model can be related to the spinodal decomposition, quark confinement, or cosmological scenario.
All predictions can be verified experimentally.
\end{abstract}

\maketitle

\section{Introduction}

The majority of physical processes are inherently nonlinear: thermal expansion, heat conductivity, lattice dynamics, nonlinear quantum optics phenomena, etc.
Some cases it may lead to phase transitions or behavior that can be interpreted as a phase transition.
Numerous works in this area indicate the importance of mentioned problems to physics.
For example, phase transition for a simple class of models was studied by Langer \cite{lang}.
Later, this method was applied in a field theory to describe the formation of new phase \cite{col}.
Other studies in this area include spinodal decomposition \cite{gor}, phase transitions with formation of a new inhomogeneous state in condensed matter \cite{lang}, alternative cosmological model \cite{LIN}, etc.

An appealing method for theoretical study of mentioned systems' class is the restatement of the problem in terms of non-linear harmonic oscillators \cite{LRZ}.
For example, stationary nuclear fission rate can be described by oscillator model with coordinate-dependent mass and specific potential \cite{Bao}.
And it turns out to be that such analytical estimates are in a good agreement with numerical simulations based on Langevine equation.
Second example comes from nuclear physics as well: transition through the fission barrier potential in the WKB-approximation \cite{Kum}.

The examples above illustrate that nonlinear oscillator models can be used in various areas of theoretical physics.
Naturally, this induces permanent interest in solving appropriate model equations.
Traditionally, behavior of nonlinear oscillators has been analyzed numerically or by perturbation methods \cite{Man}.
However, the latter cannot be correctly applied to the number of models.
For example, displacement phase transitions and dynamic soft-mode behavior require nonperturbative theories \cite{lang}.
Thus, recent advances in this area include development of alternative analytical techniques without small parameter, e.g. variational approach \cite{Man}.
Other outstanding results we should mention are the exact solution for the Hamiltonian with coordinate-dependent mass in semi-classical theory \cite{Yam} and approximations of the effective action in case of such particle moving through one-dimensional scalar field \cite{Kle}.

In the present article we use the nonlinear oscillator model to describe phase transition accompanied by formation of spatially inhomogeneous distribution of the order parameter \cite{LRZ,LZ,lub}.
We will modify the standard model of a scalar field by means of \emph{coupling} between the field and its gradient.
We expect this model to be related to spinodal decomposition \cite{gor} and cosmological scenario \cite{LIN}.
Moreover, quark confinement can be hypothetically a physical realization of this mechanistic model as well.
We assume, one can treat quarks as oscillators with increasing mass instead of specific interaction when bag's boundary is approached.
Our goal is to answer two questions: are there any analytical solutions to this model (at least under some conditions) and can this system exhibit chaotic behavior.
First question can be answered positively if certain amount of energy is contained within the system. Cases without analytic solution can be easily treated numerically.
Hereinafter we will use plots obtained by computer simulation to illustrate system's behavior at different parameters values (including the ones that lack analytic solution).

The question we state about possible chaotic behavior is not of pure curiosity.
It has been recognized that chaotic behavior seems to be preferred by some natural systems and can be utilized for practical applications \cite{Lai}.
Besides, chaotic dynamics of isolated system is interesting on its own.
For example, we know conventional measures fail to detect chaos in such systems \cite{Hab}.
Situation may become significantly different if external perturbation of Hamiltonian is employed (e.g., small external force is applied).
For instance, the nonlinear Schrodinger (NLS) equation, when perturbed, starts exhibiting  different types of chaotic behavior and instabilities (homoclinic chaos, hyperbolic resonance, and parabolic resonance).
Detailed analysis can be performed by constructing hierarchy of bifurcations \cite{Shl}.
Besides, numerical computation show that NLS driven by external force is subject to similar chaotic phenomena \cite{Shl}.
This suggests the idea that our model system can undergo bifurcation or acquire chaotic properties as well.

\section{Model with order parameter and its gradient coupling}

Let us consider a continuous system.
In the context of phase transitions theory, we assume it to have the ground state described in terms of order parameter.
The latter can be the subject to various geometrical representations.
For instance, theories of condensed matter \cite{lang} or field theories \cite{LIN} introduce order parameter in form of scalar field.

Determining a stable state of a condensed matter, we can expect a non-uniform field distribution in some cases.
And indeed, current experimental data suggests the existence of disordered configurations of the ground state.
Their theoretical descriptions are mostly phenomenological, but fairly general.
For example, \cite{LEV1} and \cite{ZAG} describe spatial distributions of the order parameter both before and after the phase transition.

The model can be further improved if a term responsible for interaction between the order parameter and its gradient is introduced into the free-energy functional.
In case of competing order parameters, the gradient terms may lead to inhomogeneous states \cite{Nus}.
Moreover, at instability threshold coefficients before the quadratic terms of the order parameter and its gradient can change their signs \cite{gor}.
This means, some decomposition scenario is possible and system's state may undergo transition from disordered to modulated, patterned, or ordered-patterned.
In other words, we can observe spinodal decomposition when scalar spatially-dependent order parameter rapidly changes.
We expect this behavior to be related to cosmological model.
For example, it can describe generation of spatially inhomogeneous states when temperature drops.
Besides, temperature decrease can be the reason for a new bubble phase formation in the cosmological model.

The main idea of present contribution is to generalize the standard model of scalar field by introducing the interaction between the field and its gradient.
%This generalization is reasonable in view of both mathematical and physical arguments.
We consider following form of the condensed matter free energy density \cite{lang}
\begin{equation}
\label{eq_free_energy}
\begin{aligned}
     f&= a \left(\vec{\nabla} \varphi(\vec{r}\,) \right)^{2}
       + b \left(1 - \left(\varphi(\vec{r}\,)\right)^{2}\right)^{2} +\\
      &+ c \varphi^{2}(\vec{r}\,)\left(\vec{\nabla}\varphi(\vec{r}\,)\right)^{2},
\end{aligned}
\end{equation}
where $a$, $b$, and $c$ are real constants ($a, b > 0$).
The same expression is valid for three-dimensional action in field theory \cite{LIN}.
One can see, setting $c = 0$ transforms \eqref{eq_free_energy} to accustomed relation of the $\varphi^{4}$ \emph{model} for single real scalar field.

% Це для випадку с = 0, чи для довільного?
%In the condensed matter physics we have presentation of the free energy after the phase transition when the temperature of a system $T$ is less than its critical value $T_c$.

There are two reasons why form \eqref{eq_free_energy} is favored over other possible examples.
First of all, we can prove this model is equivalent to the mechanical one\,--- nonlinear oscillator with coordinate-dependent mass.
It seems, renormalization problem can be solved in this new representation by means of QFT methods.
Namely, one can imply covariant background field method \cite{Kle}, to calculate the one-loop quantum effective action for the particle with the coordinate-dependent mass moving slowly trough one-dimensional configuration space.
This makes \eqref{eq_free_energy} quite a promising model, because thus far we have been working with free energy expressed in terms of non-renormalizable fields, masses, and coupling constants.

The second reason we propose model \eqref{eq_free_energy} is the variety of its solutions.
We will describe this solutions and show they can give rise to spatially inhomogeneous scalar field distributions and topological structures of the new phase.

\section{Order parameter-space\,--- space-time duality}

Let us define free energy from its density \eqref{eq_free_energy} as follows
$F[\varphi(\vec{r}\,)] = \iiint f\left(\varphi(\vec{r}\,), \vec{\nabla} \varphi(\vec{r}\,) \right) d\vec{r}$,
where integration is performed over the entire system.
We write $F[\varphi(\vec{r}\,)]$ here to emphasize $F$ is treated as functional of $\varphi$ (and a subject to variational calculus we will apply).

Now we can perform minimization of $F[\varphi(\vec{r}\,)]$ by means of variational calculus, namely calculate functional derivative $\delta F / \delta \varphi(\vec{r}\,)$.
This is a multidimensional problem and its solution can be found by following Euler-Lagrange-Ostrogradsky equation
$$
      \frac{\partial f}{\partial \varphi}
    - \sum_i \frac{\partial}{\partial r_i} \frac{\partial f}{\partial\, (\partial \varphi/\partial r_i)}
    = 0,
$$
where $r_i$ are components of $\vec{r}$.
Before we proceed, let us rewrite the last equation in a different form
\begin{equation}
\label{eq_var_deriv}
      \frac{\partial f}{\partial \varphi} 
    - \vec{\nabla} \cdot \frac{\partial f}{\partial \vec{\nabla} \varphi}
    = 0,
\end{equation}
where $\partial f/\partial \vec{\nabla} \varphi$ is a derivative of a scalar with respect to a vector and $\cdot$ means scalar product.

If $\vec{\nabla}$ in equation \eqref{eq_var_deriv} was one-dimensional derivative, we could call that dimension ``time'' and designate it $t$, while $F$ could be renamed to $L$ and called Lagrangian.
If so, we could obtain a regular equation we used to see in classical mechanics.
Actually, this is exactly what will happen in a moment, after we apply symmetry and reduce the number of spatial variables to one.

Now we use equations \eqref{eq_free_energy} and \eqref{eq_var_deriv} to obtain the following expression
\begin{multline}
\label{eq_eu_lag_vect}
      \left(a + c \left(\varphi(\vec{r}\,)\right)^2 \right) \Delta\varphi(\vec{r}\,)
    + c \varphi(\vec{r}\,) \left(\vec{\nabla}\varphi(\vec{r}\,)\right)^2 +\\
    + 2 b \varphi(\vec{r}\,) \left(1 - \left(\varphi(\vec{r}\,)\right)^2\right)
    = 0.
\end{multline}
We are interested in bubble formation, thus spherical symmetry can be applied.
For one- and two-dimensional cases this means we can write $d\varphi / dr$ instead of $\vec{\nabla}\varphi$ and $d^2\varphi / dr^2$ in place of $\Delta\varphi$.
Three-dimensional case is a little bit more complicated, since term $(2/r) d\varphi / dr$ appears from Laplace operator.
But in this case one can imply a thin-wall approximation \cite{LIN,col}, which essentially means neglecting $(2/r) d\varphi / dr$ term.
Thus, we write for all one to three dimensions
\begin{multline}
\label{eq_eu_lag_symm}
      \left(a + c \left(\varphi(r)\right)^2 \right) \frac{d^2 \varphi}{dr^2}
    + c \varphi(r) \left(\frac{d \varphi}{dr}\right)^2 +\\
    + 2 b \varphi(r) \left(1 - \left(\varphi(r)\right)^2\right)
    = 0,
\end{multline}
where $r$ designates the only coordinate left.
%In the case of different dimensions for the bubble formation of a new phase it is necessary to determine the surface energy of the bubble wall.

Now we leave cosmological model for a moment and switch gears to the mechanical one\,--- oscillator with coordinate-dependent mass.

\begin{figure}[ht]
\begin{center}
\includegraphics[width=0.6\columnwidth]{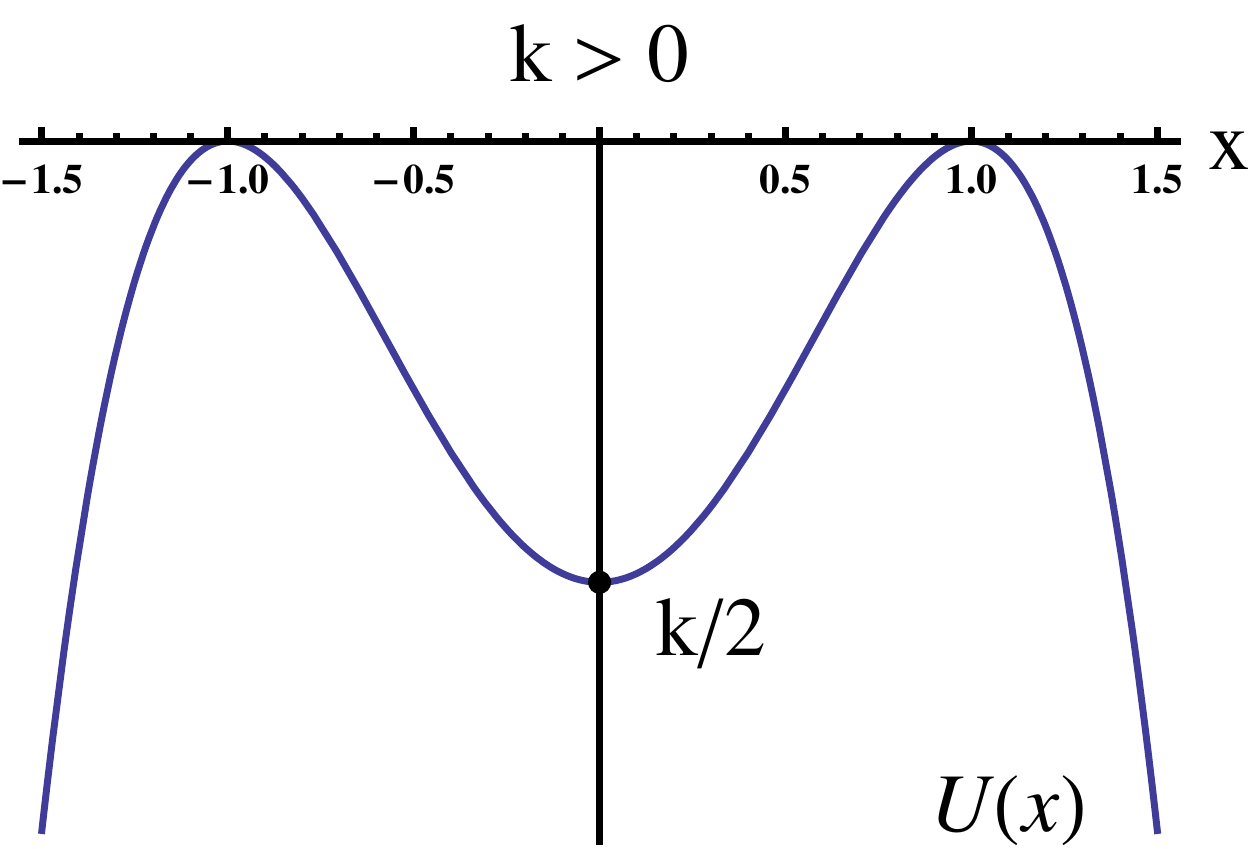}
\end{center}
\caption{Potential energy.}
\label{fig_potential}
\end{figure}

We start with a model Lagrangian \cite{LRZ,LZ}
\begin{equation}
\label{eq_lagrangian}
L = \frac{\left(a + c x^2\right) \dot{x}^2}{2}
  - \frac{-b \left(x^2 - 1\right)^2}{2},
\end{equation}
where $\dot{x}=dx/dt$ denotes velocity.
We write $-b$ to keep $b$ positive due to its physical meaning as elasticity coefficient.
Besides, one should notice, current Lagrangian describes particle of a variable coordinate-dependent mass $m_{\text{total}} = a + c x^2$.
Potential affecting this particle is shown in figure~\ref{fig_potential}.

From presented Lagrangian we can write Euler-Lagrange equation for this system
\begin{multline}
\label{eq_eu_lag_1}
      \left(a + c \left(x(t)\right)^2 \right) \frac{d^2 x}{dt^2}
    + c x(t) \left(\frac{dx}{dt}\right)^2 +\\
    + 2 b x(t) \left(1 - \left(x(t)\right)^2\right)
    = 0.
\end{multline}
Now we compare equations \eqref{eq_eu_lag_1} and \eqref{eq_eu_lag_symm} to see they are essentially the same.
By this comparison we establish duality between order parameter $\varphi$ and spatial coordinate $x$ ($\varphi \leftrightarrow x$) as well as between spatial coordinate $r$ and time $t$ ($r \leftrightarrow t$).
Presumably, this duality between thermodynamical and mechanical systems may hold in some different models as well.

In the next section we will consider equations for oscillator with coordinate-dependent mass more carefully and present possible analytical solutions.
But before one more remark should be done.
If one states the problem of homogeneous free energy distribution (e.g. $\vec{\nabla}f = 0$, $f$ from \eqref{eq_free_energy} being used), but claims order parameter is not a constant (e.g. $\vec{\nabla}\varphi \neq 0$), he will get equation almost similar to \eqref{eq_eu_lag_vect}
\begin{multline*}
      \left(a + c \left(\varphi(\vec{r}\,)\right)^2 \right) \Delta\varphi(\vec{r}\,)
    + c \varphi(\vec{r}\,) \left(\vec{\nabla}\varphi(\vec{r}\,)\right)^2 -\\
    - 2 b \varphi(\vec{r}\,) \left(1 - \left(\varphi(\vec{r}\,)\right)^2\right)
    = 0.
\end{multline*}
The only difference seen is sign in front of $b$.
This equation can be described with the same mechanical model \eqref{eq_eu_lag_1} just letting coefficient $b$ to be negative.
We do not expect this case to have much physical implication, but we will briefly mention it at the end of the next section.

\section{Non-linear oscillator with coordinate-dependent mass}

At this point we have already established connection between physical model and nonlinear oscillator.
Now we will perform analysis of the oscillator (actually we do that twice, for $b > 0$ and $b < 0$).
First of all, we will calculate analytical equation for phase curves.
Some of phase trajectories will appear to be closed, thus corresponding to periodic motion.
So we will find approximate expression for this period.
Than we present analytical solutions for some values of parameters and formula that greatly simplifies numerical calculation.
We finish analysis by applying small external driving force to the oscillator and exploring the possibility of chaotic behavior by means of Poincare sections and Lyapunov characteristic exponents.

Equation \eqref{eq_eu_lag_1} contains three parameters ($a$, $b$, and $c$) that is a little bit inconvenient.
Thus, we perform substitution
$$
             x(t) = x(\tau(t)),
    \quad \tau(t) = \sqrt{\frac{2 b}{a}}\, t,
    \quad \alpha  = \frac{c}{a}.
$$
We do not use absolute value in radicand since $a > 0$ and $b > 0$.
Simplification yields
\begin{equation}
\label{eq_eu_lag}
    \left(1 + \alpha x^2\right) \ddot{x} + \alpha x \dot{x}^2 + x \left(1 - x^2\right) = 0,
\end{equation}
where $\dot{x}$ now denotes the derivative with respect to $\tau$, not $t$ as previously: $\dot{x} = dx/d\tau$.
Now we have the only parameter $\alpha$ left that seems quite convenient.

Before we proceed further, let's outline some features of \eqref{eq_eu_lag}.
First of all, if $x(\tau)$ is a solution, $-x(\tau)$ is solution as well.
We will use this fact throughout computation without special emphasis.
Second observation has more to do with physics of the system.
Expression for the only parameter $\alpha$ does not contain $b$.
Thus, phase portrait should not depend on $b$.
In other words, $b$ does not change the phase curve, but only the speed system is traversing along.

Expression \eqref{eq_eu_lag} can be reduced to the Bernoulli equation by following substitution
$$
    u(x) = \dot{x}^2 \Rightarrow 2\dot{x}\ddot{x} = \frac{d}{d\tau} \left(\dot{x}^2\right)
                                                  = \frac{du}{d\tau}
                                                  = \frac{du}{dx} \frac{dx}{d\tau}
                                                  = \dot{x} \frac{du}{dx},
$$
that yields equation
$$
    \frac{du}{dx} + \frac{2 \alpha x}{1 + \alpha x^2}\, u + \frac{2 x (1 - x^2)}{1 + \alpha x^2} = 0,
$$
which can be integrated.
As result, we obtain following relation
$$
     u(x) = \frac{\left(1 - x^2\right)^2 + C}{2(1 + \alpha x^2)},
$$
where $C$ is integration constant.

Constant $C$ in the last equation seems to have a clear physical meaning, we want to unleash.
First of all, we rewrite the last equality as follows
\begin{equation}
\label{eq_first_d}
    2 \left(1 + \alpha x^2\right) \dot{x}^2 - \left(1 - x^2\right)^2 = C.
\end{equation}
Now let's focus on \eqref{eq_lagrangian}.
One may notice, it does not contain any explicit dependence on time, $\partial L/\partial t = 0$.
Thus, we expect the total energy to be conserved within the system
$$
E = \frac{\left(a + c x^2\right) \dot{x}^2}{2}
  + \frac{-b \left(x^2 - 1\right)^2}{2},
$$
where $E$ is total energy (we do not designate it $H$, Hamiltonian, only to emphasize it is constant).
Now, performing substitution $x(t) \rightarrow x(\tau)$ and few algebraic manipulation, we get
\begin{equation}
\label{eq_first_int}
    \varepsilon = \frac{2 E}{b}
                = 2 \left(1 + \alpha x^2\right) \dot{x}^2 - \left(1 - x^2\right)^2,
\end{equation}
where $\varepsilon$ is dimensionless energy.
Now we clearly see connection between \eqref{eq_first_d} and \eqref{eq_first_int} and thus physical meaning of $C$.

\begin{figure}[ht]
\begin{center}
\includegraphics[width=\columnwidth]{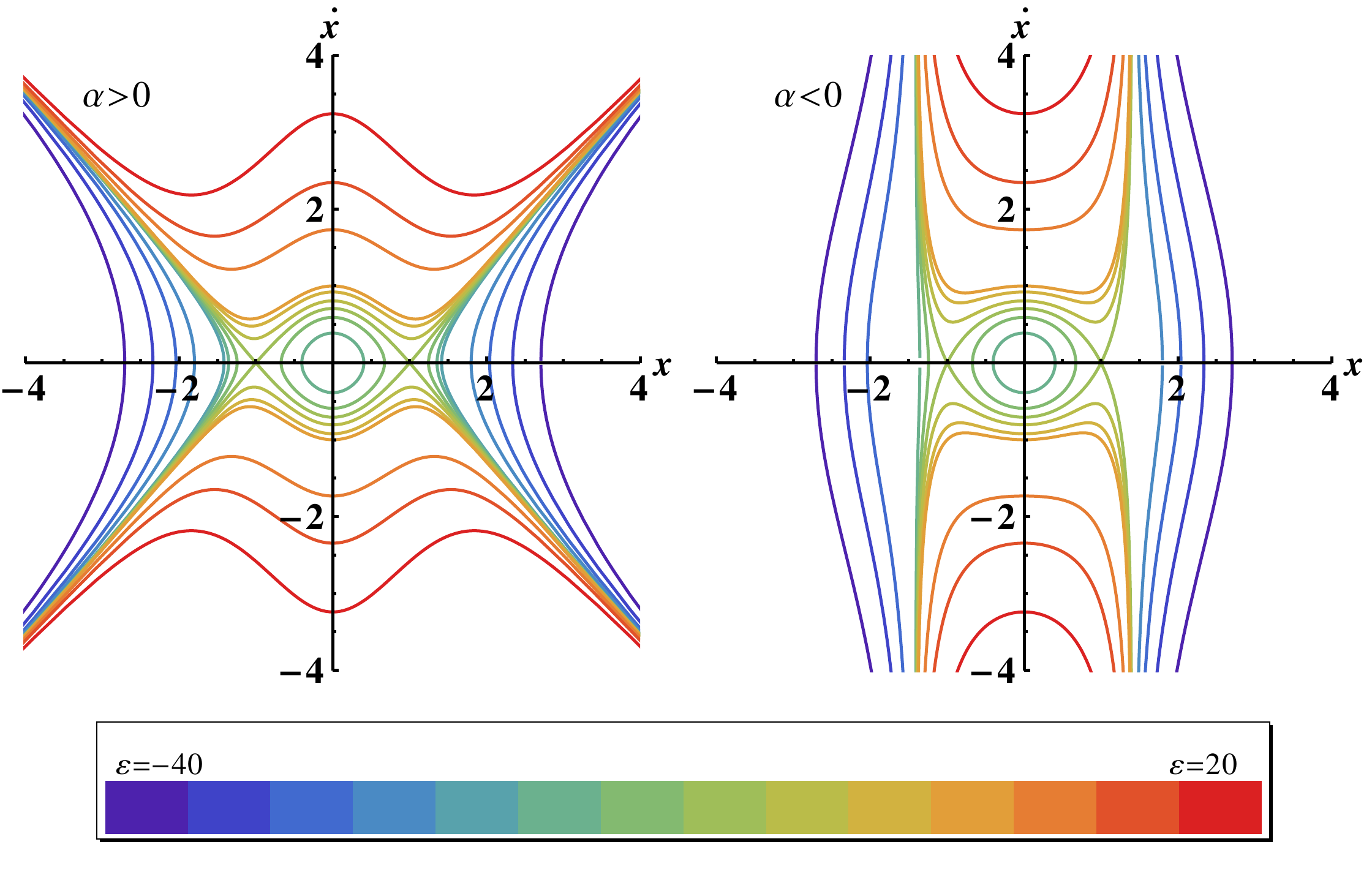}
\end{center}
\caption{Phase portraits depending on total energy of the system $\varepsilon$. $\alpha = \pm 0.5$.}
\label{fig_phase_portr}
\end{figure}

We can rewrite \eqref{eq_first_int} as follows
\begin{equation}
\label{eq_first_int1}
    \dot{x}^2 = \frac{\left(1 - x^2\right)^2 + \varepsilon}{2(1 + \alpha x^2)},
\end{equation}
to perform further analysis.
First of all, \eqref{eq_first_int1} provides us with a convenient tool to draw the phase portrait of the system.
Basically, we can simply choose some value for energy $\varepsilon$ and plot $\dot{x}$ as function of $x$.
Result is presented in figure \ref{fig_phase_portr}.

One can see from figure \ref{fig_phase_portr}, there exist bounded and unbounded solutions depending on parameters and initial conditions.
Let us focus for a moment on bounded solutions.
In these cases we expect the oscillator to move back and forth in a periodic manner.
And we can approximately find its vibration period.

There are plenty of books and papers devoted to the problem of strongly nonlinear oscillator's limit cycle.
At least \cite{Del,And,Ben,Nay,He1,He3} should be mentioned, but that's certainly not the full list.
We will generally follow guidelines of \cite{Nay} when performing our solution.

First of all, if $x(\tau)$ is periodic we expect it to be expandable in Fourier series
$$
    x(\tau) = \frac{A_0}{2} + \sum_{n=1}^{\infty} A_n \sin\left(\frac{2\pi n \tau}{T} + \varphi_n\right),
$$
where $T$ is the period.
From \eqref{eq_first_int1} we see if pair $\{x(\tau),\dot{x}(\tau)\}$ is solution than $\{-x(\tau),-\dot{x}(\tau)\}$ is solution as well.
This means during one period along with every point $x$ particle will inevitably ``visit'' point $-x$.
As result we expect
$$
    \int\limits_{\tau_0}^{\tau_0+T} x(\tau) d\tau = 0,
$$
where $\tau_0$ is arbitrary.
Thus, we conclude $A_0 = 0$.

Now we assume the largest coefficient $A$ is multiplied by sine having period equal to the period of $x(\tau)$.
In other words we approximate
$$
    x(\tau) \approx A \sin\left(\frac{2\pi\tau}{T} + \varphi\right).
$$
Applying this assumption to \eqref{eq_first_int} we get
\begin{multline*}
   - A^4 \sin^4(\omega\tau + \varphi) \left(1 + 2\alpha\omega^2\right) +\\
   + 2 A^2 \sin^2(\omega\tau + \varphi) \left(1 - \omega^2 + \alpha A^2 \omega^2\right) +\\
   + \left(2 A^2 \omega^2 - \varepsilon - 1\right)
   = 0,
\end{multline*}
where $\omega = 2\pi/T$.
If one recalls potential influencing the particle (figure \ref{fig_potential}), it is naturally to expect $|A| < 1$.
Thus, we neglect term with $A^4\sin^4$ and claim the equation should hold for every $\tau$.
This yields two equations
$$
\left\{
    \begin{aligned}
        1 - \omega^2 + \alpha A^2 \omega^2 &= 0,\\
        2 A^2 \omega^2 - \varepsilon - 1   &= 0.
    \end{aligned}
\right.
$$
Solving this system and converting $\omega$ to $T$ we finally get
\begin{equation}
\label{eq_period}
    T = \frac{2\sqrt{2}\pi}{\sqrt{2 + \alpha(\varepsilon + 1)}}.
\end{equation}
For negative $\alpha$ period $T$ lengthens if energy $\varepsilon$ is increased.
But there is inverse situation for positive $\alpha$: period shortens if we increase the energy $\varepsilon$.
If $\alpha = 0$ period does not depend on energy $\varepsilon$ at all.

One should notice, expression \eqref{eq_period} does not exist for every possible value of $\varepsilon$ and $\alpha$.
This is quite naturally: as is seen from figure \ref{fig_phase_portr}, not every combination of energy and $\alpha$ is suitable for periodic motion.

At this point we know the existence of two types of solutions: bounded and unbounded.
Besides, we did analysis of bounded solutions assuming they are periodic.
We leave the problem of phase curves stability and possible stochastization until the end of this section.
And for now we proceed the analysis of \eqref{eq_first_int1}.

Formally, we can write solution of \eqref{eq_eu_lag} by taking square root and integrating \eqref{eq_first_int1}
\begin{equation}
\label{eq_formal_solution}
     \tau =
     \int\limits_{x_0}^{x}
         \sqrt{
             \frac{2 \left(1 + \alpha x^2\right)}
                  {\left(1 - x^2\right)^2 + \varepsilon}
          }\, dx.
\end{equation}
One can use \eqref{eq_formal_solution} to obtain numerical solution of the problem under consideration.
There is one precaution we should do.
Depending on parameter $\varepsilon$ value integrand can be defined on all real line or on three non-intersecting subsets.
This fact has a clear physical meaning: if particle lacks energy to ``climb the hills'' $-1$ and $+1$ (figure \ref{fig_potential}), ``forbidden'' zones ($x$ values) appear.
Thus, we will have three regions: to the left of $-1$, to the right of $+1$ and one in-between.
Surely, $x_0$ and $x$ should be in the same region.
We suppose they are and the one performing integration took care of choosing reasonable integration limits.

Nevertheless, some parameter values allow not only numerical but analytic solution as well.
Even better, this solutions seem to have physical meaning.
So let us introduce few possible cases.

Obviously, the easiest case we can think of is the one with zero parameters, $\varepsilon = 0$, $\alpha = 0$.
Equation \eqref{eq_formal_solution} reduces to
$$
     \tau = 
     \int\limits_{x_0}^{x} 
             \frac{\sqrt{2}}{\left|1 - x^2\right|} \, dx.
$$
We can easily perform integration and find the inverse function
\begin{equation}
\label{eq_cosmo}
    x = \pm\tanh\left(\frac{\tau + \tau_0}{\sqrt{2}}\right),
\end{equation}
where $\tau_0 = \sqrt{2}\tanh^{-1}(x_0)$.
Sign is positive if $|x| < 1$ and negative otherwise.
As one can see, \eqref{eq_cosmo}, so-called kink solution, fully recovers the behavior of the fundamental scalar field in the standard cosmological model \cite{LIN}.

One more partial solution can be obtained if $\alpha = -1$ and $\varepsilon = 0$.
Equation \eqref{eq_formal_solution} reduces to
$$
     \tau = 
     \int\limits_{x_0}^{x} 
         \sqrt{
             \frac{2}{1 - x^2}
          }\, dx.
$$
Integration is easily performed as well to yield
\begin{equation}
\label{eq_sine}
    x = \sin\left(\frac{\tau + \tau_0}{\sqrt{2}}\right),
\end{equation}
where $\tau_0 = \sqrt{2}\sin^{-1}(x_0)$.
As result we get periodic solution.
One can check its period $2\sqrt{2}\pi$ is consistent with result we get from equation \eqref{eq_period}.
Notable, integration can be performed only if $-1 \leq x \leq 1$ during integration.
This makes sense, since, as we know from figure \ref{fig_phase_portr}, periodic trajectories are close to zero and do not outspread certain, dependent on parameters $\alpha$ and $\varepsilon$, distance.

One more case we want to analyze is $\varepsilon = 0$.
We can perform integration of \eqref{eq_formal_solution}
\begin{multline}
\label{eq_e_0_solution}
    \tau + \tau_0
  = \pm\Biggl[\sqrt{1 + \alpha} \sinh^{-1} \left(x \sqrt{\frac{1 + \alpha}{1 - x^2}}\right) -\Biggr.\\
  - \Biggl.\sqrt{\alpha} \sinh^{-1} \left(x \sqrt{\alpha}\right)\Biggr],
\end{multline}
where sign $\pm$ is taken positive if $|x| < 1$ and negative otherwise.
With respect to this solution, one more detail should be mentioned.
Depending on $\alpha$ value $\sqrt{\alpha}$ and $\sqrt{1 + \alpha}$ can be either real or imaginary.
We can write out three different versions of \eqref{eq_e_0_solution}, but for the sake of brevity we suppose one uses equation $\sinh^{-1}(i x) = i \sin^{-1}(x)$ to generate all three cases.

\begin{figure}[ht]
\begin{center}
\includegraphics[width=\columnwidth]{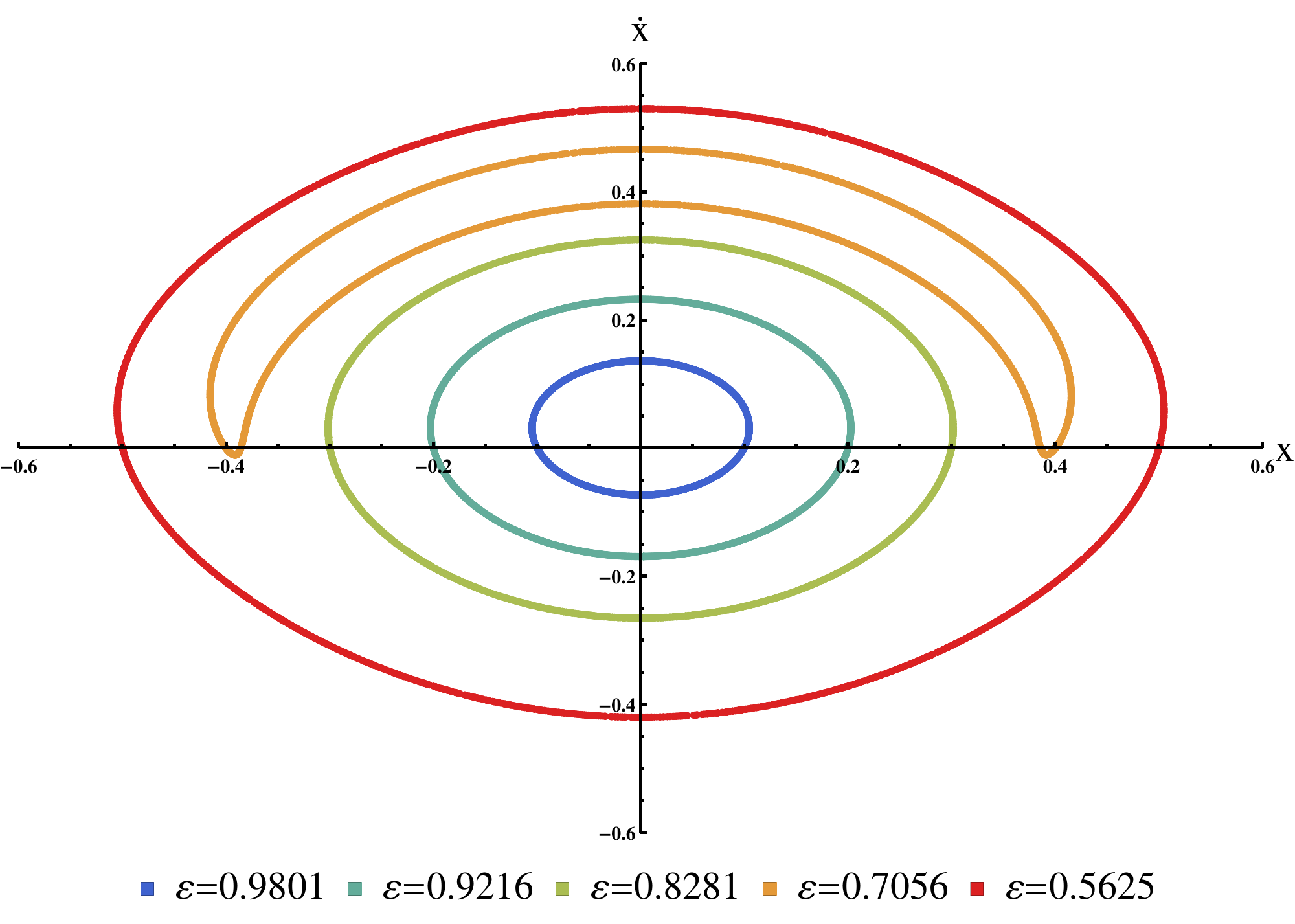}
\end{center}
\caption{Poincare section (also known as Poincare map or first recurrence map). $\alpha = 0.5$, $\omega = 0.3$, $A = 0.1$.}
\label{fig_poincare}
\end{figure}

Now let us consider the case: oscillator driven by periodic external force
\begin{equation}
\label{eq_driven}
    \left(1 + \alpha x^2\right) \ddot{x} + \alpha x \dot{x}^2 + x \left(1 - x^2\right) = A \sin(\omega \tau),
\end{equation}
where $A$ and $\omega$ are amplitude and frequency of the force.
We use numerical simulation to explore stability of the system by means of Poincare sections (figure \ref{fig_poincare}).
It seems, phase trajectories are quasiperiodic and stable.

\begin{figure}[ht]
\begin{center}
\includegraphics[width=0.6\columnwidth]{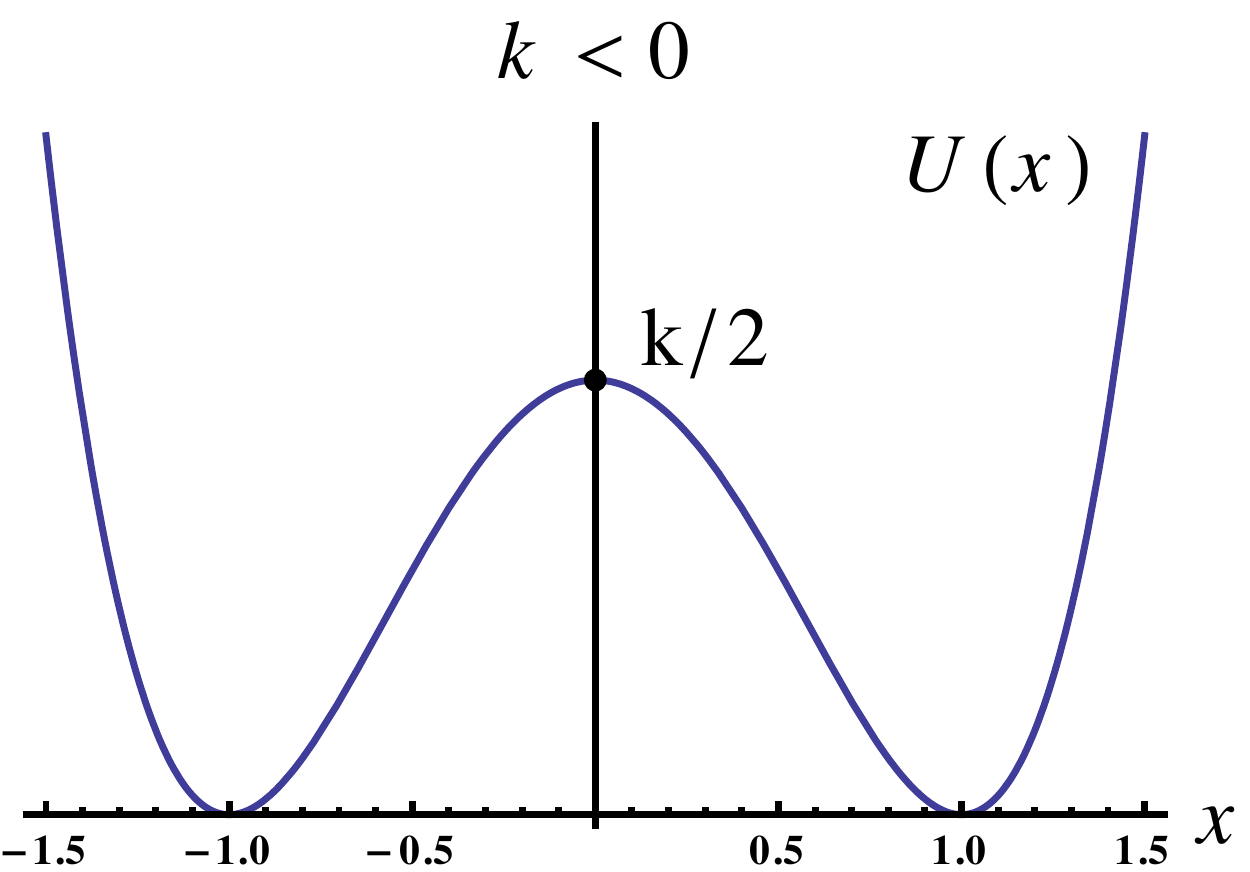}
\end{center}
\caption{Potential energy.}
\label{fig_potential_minus}
\end{figure}

System behavior crucially changes if we allow $b$ (equation \eqref{eq_lagrangian}) to be negative.
As is seen from figure \ref{fig_potential_minus}, negative $b$ changes potential crucially.
Thus, we expect system behavior to be significantly different as well.

First, looking at potential in the figure \ref{fig_potential_minus}, we can expect particle to move periodically.
When energy is low, particle should oscillate in the neighborhood of $-1$ or $+1$.
Granted with more energy, it should move back and forth visiting both $+1$ and $-1$ potential wells.
As we will see, this is exactly what happens when $\alpha > 0$ and just slightly different is situation for $\alpha < 0$.

\begin{figure}[ht]
\begin{center}
\includegraphics[width=\columnwidth]{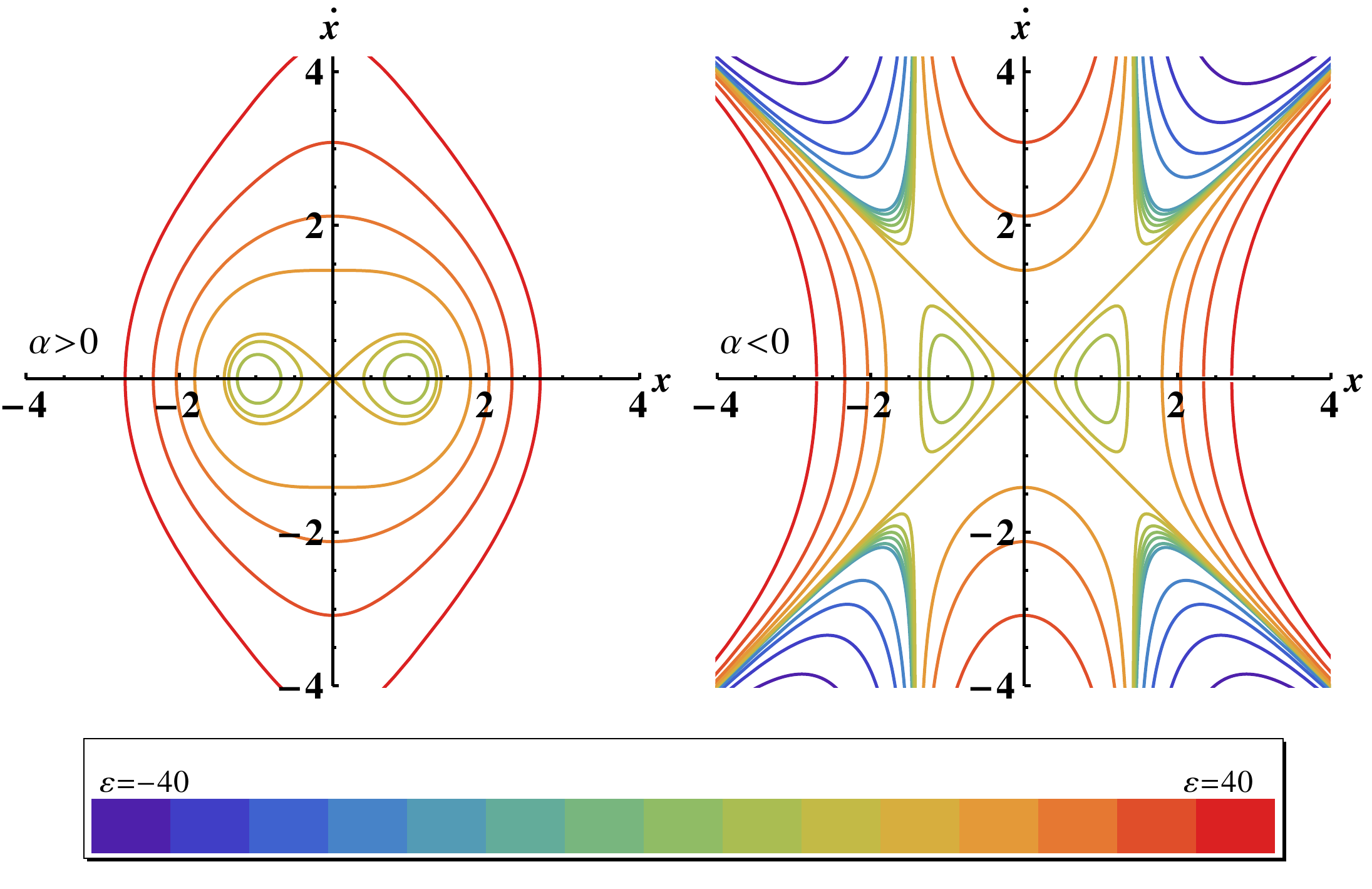}
\end{center}
\caption{Phase portraits depending on total energy of the system $\varepsilon$. $\alpha = \pm 0.5$.}
\label{fig_phase_portr_minus}
\end{figure}

Now let us look what happens to our calculations if we suppose $b < 0$ in \eqref{eq_lagrangian}.
First difference appears when substitution $t \rightarrow \tau$ is performed.
Since $b$ is negative we should consider $\tau(t) = \sqrt{|2b/m|}\,t$ and \eqref{eq_eu_lag} changes to
\begin{equation}
\label{eq_eu_lag_minus}
    \left(1 + \alpha x^2\right) \ddot{x} + \alpha x \dot{x}^2 - x \left(1 - x^2\right) = 0,
\end{equation}
where minus sign in front of $x \left(1 - x^2\right)$ is the only noticeable difference.
This is result of $b$ by $|b|$ division, since the former is negative.
Repeating all substitutions and first integration we get equivalent of \eqref{eq_first_int}
\begin{equation}
\label{eq_first_int_minus}
    \varepsilon = -\frac{2 E}{b}
                = 2 \left(1 + \alpha x^2\right) \dot{x}^2 + \left(1 - x^2\right)^2.
\end{equation}
Equation \eqref{eq_first_int_minus} can be used to draw the phase portrait, figure \ref{fig_phase_portr_minus}.

\begin{figure}[ht]
\begin{center}
\includegraphics[width=\columnwidth]{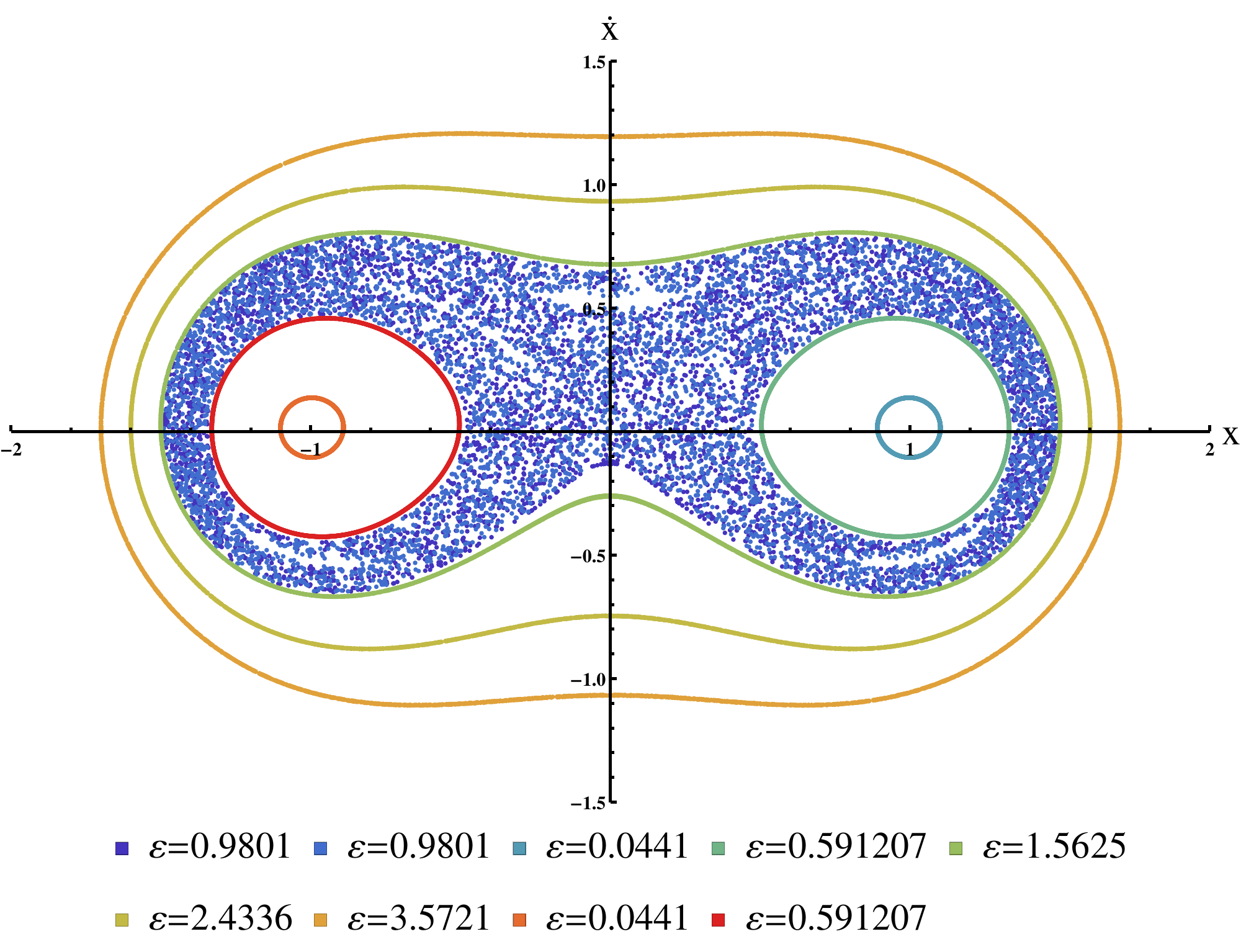}
\end{center}
\caption{Poincare section, $\alpha = 0.5$, $\omega = 0.3$, $A = 0.1$.
         Stochastization is clearly seen for one of $\varepsilon$'s values.}
\label{fig_poincare_minus}
\end{figure}

Now let us consider stability of periodic solutions.
We expect both stable and unstable trajectories to be seen.
In other words, we predict chaos in this system for certain values of particle's energy and amplitude of external perturbation.
From physical point of view we expect the following.
Suppose, particle has just enough energy to get close to local potential maximum at zero (figure \ref{fig_potential_minus}) but not to ``jump it over''.
This moment external perturbation can be sufficient to ``leap over'' the particle that otherwise would be periodically oscillating within one potential well.
Such significant change (unexpected jump from one potential well to another) should destroy periodic movement pattern and we will observe chaos.
To justify this rather general idea we turn to numerical computations and draw Poincare sections (figure \ref{fig_poincare_minus}).

To be even more convinced figure \ref{fig_poincare_minus} represents chaos, we used module for Wolfram Mathematica \cite{San} to obtain Lyapunov exponents.
For ``chaotic region'' in figure \ref{fig_poincare_minus} there are three Lyapunov exponents: greater than zero, equal to, and less than zero.
According to classification in \cite{San} this is $C^1$ chaos.
%The non-linear oscillator model for phase transition accompanied by appearance of spatially inhomogeneous distribution of the order parameter was proposed.
%We explored its behavior with different values of the parameters: the ones that can be interpreted as pre-phase-transitional $b < 0$ and post-phase-transitional $b > 0$.

\section{Results and discussion}

In this part we will briefly recall pivoting equations from the previous section and describe their physical meaning in the light of duality we have established ($\varphi \leftrightarrow x$, $r \leftrightarrow t$).

First, we have obtained the energy of the oscillator \eqref{eq_first_int}.
In terms of mechanical model this result was valuable because it leads to \eqref{eq_first_int1}\,--- analytical expression for all phase curves.
Analyzing Lagrangian, we showed that energy $\varepsilon$ is conserved within this model.
Together with $\alpha$ it defines the phase curve system will traverse along.
Using established duality $\varphi \leftrightarrow x$, $r \leftrightarrow t$ on \eqref{eq_first_int}, one can obtain invariant in terms of order parameter $\varphi$ and distance $r$.
Thus, measuring order parameter and its derivative at few different spatial positions, one can experimentally verify presented theory by means of \eqref{eq_first_int}.

Some of phase trajectories appear to be closed (figure \ref{fig_phase_portr}), thus indicating periodic motion.
The period of corresponding cycle was found to be approximated by \eqref{eq_period}.
In terms of order parameter and space this means we have found possible periodic or quasiperiodic distribution and its characteristic length.

Now let us return to the problem of finding \eqref{eq_eu_lag}'s solution.
Some formal expression is established by \eqref{eq_formal_solution}, but when it comes to numerical simulation this equation becomes really valuable.
Few following equations \eqref{eq_cosmo}, \eqref{eq_sine}, and \eqref{eq_e_0_solution} are analytical solutions to the stated problem at certain values of parameters.
Particularly, one of them recovers the behavior of the fundamental scalar field in the standard cosmological model.
Generally speaking, we have obtained different solution families: kink, periodic, quasiperiodic, etc.
Notably, periodic solutions do not exist in any standard model of phase transition \cite{lang} or scalar field \cite{LIN}.

Plenty qualitatively different solutions offer the possibility of new phase transition scenario.
Let us suppose, system behaves according to \eqref{eq_free_energy}, but its coefficients, or at least $c$, are very slowly changing in time.
If this is the case, we may expect system to evolve by slowly transitioning from one solution of \eqref{eq_eu_lag} (actually, its dual in terms of order parameter-distance) to another.
But occasionally solution of \eqref{eq_free_energy} may crucially change in response to small $c$ variation.
As an example, one may check what happens to \eqref{eq_e_0_solution} if $\alpha$ changes its sign or drops below $-1$ ($\alpha$ is proportional to $c$).
If it happens, we will observe a new type of phase transition originated from coupling between the order parameter and its gradient.

Regarding physical implications, there are two questions to be answered.
First, why does $c$ change at all?
In current model we postulated the form of free energy, so there is no way to derive change of $c$ within its scope.
But presumably, it may be caused by external parameters: electromagnetic field, pressure, temperature, etc.
Critical values of this parameters($T_c$, $E_c$, etc.) should be associated with critical values of $\alpha$ (proportional to $c$) that lead to crucial changes in solutions of \eqref{eq_free_energy}.

The other question is about physical systems, it can be applicable to.
We expect predicted effects to be seen in spinodal decomposition experiments with decreasing temperature \cite{gor} or within standard cosmological model, when bubble formation and evolution are studied \cite{LIN} (for topological point of view please see \cite{Baz}).
Particularly, formation of the new phase bubble corresponds to the limit state of the spatially-periodical order parameter (scalar field) \cite{lang}.

And last, but not the least\,--- case $b < 0$.
It was presented in less details, but all methods used on $b > 0$ were perfectly applicable.
Solutions for $b < 0$ are interesting, because of their behavior under external perturbation.
We did not found chaotic behavior for $b > 0$ (figure \ref{fig_poincare}), by Poincare sections and numerical calculation of Lyapunov exponents for $b < 0$ undoubtedly indicate chaos in this system (figure \ref{fig_poincare_minus}).
This means, systems with uniform free energy distribution (constant free energy density) have either periodic or chaotic distribution of the order parameter.
Presumably, this stochastization can be verified experimentally.

It is worth noting that system's behavior crucially changes when coupling constant switches its sign to negative.
Both for $b > 0$ and $b < 0$ we present phase diagrams for $\alpha > 0$ and $\alpha < 0$ side-by-side to make the difference more evident (figures \ref{fig_phase_portr} and \ref{fig_phase_portr_minus}).
Most interesting is the case $b > 0$ and $\alpha < 0$.
When being between $-1$ and $+1$, now matter how large the $\varepsilon$ is, particle can't escape this region.
We suppose, this may be a model for quarks as they cannot escape the bag.

\section{Conclusions}

To conclude, we have considered a model of the first order phase transitions, modified by introduction of coupling between order parameter and its gradient.
An equivalent mechanical model was found and studied.
By means of this model, we have discovered certain invariant in terms of order parameter $\varphi$ and distance $r$.
Analyzing solutions to the model, both analytically and numerically, we showed the possibility of the \emph{new} scenario of phase transition.
Considered as a byproduct, systems with evenly distributed free energy were studied within the framework of mentioned mechanical model.
Worth noting is stochastization we have obtained at certain values of parameters.
We expect our results to be applicable to number of physical systems: cosmological model, spinodal decomposition, quark confinement, etc. and experimentally verifiable at least in some cases.

\end{document}